\documentclass[pra,reprint,twocolumn]{revtex4}
\usepackage{amsthm}
\usepackage{amsmath}
\usepackage{amsfonts}
\usepackage{graphicx}
\usepackage{dcolumn}
\usepackage{mathrsfs}
\usepackage{float}
\usepackage{multirow}

\newcommand{\ket}[1]{|#1\rangle}
\newcommand{\bra}[1]{\langle#1|}

\begin{document}

\title{Fixed-point quantum search with an optimal number of queries}

\author{Theodore J. Yoder, Guang Hao Low, Isaac L. Chuang}
\affiliation{Massachusetts Institute of Technology}
\date{\today}

\begin{abstract}
Grover's quantum search and its generalization, quantum amplitude amplification, provide quadratic advantage over classical algorithms for a diverse set of tasks, but are tricky to use without knowing beforehand what fraction $\lambda$ of the initial state is comprised of the target states. In contrast, fixed-point search algorithms need only a reliable lower bound on this fraction, but, as a consequence, lose the very quadratic advantage that makes Grover's algorithm so appealing. Here we provide the first version of amplitude amplification that achieves fixed-point behavior without sacrificing the quantum speedup. Our result incorporates an adjustable bound on the failure probability, and, for a given number of oracle queries, guarantees that this bound is satisfied over the broadest possible range of $\lambda$.
\end{abstract}

\maketitle

Grover's quantum search algorithm \cite{Grover1996} provides a quadratic speedup over classical algorithms for solving a broad class of problems. Included are the many important, yet computationally prohibitive NP problems \cite{Bennett1997}, for which finding a solution reduces to searching for one. Because the problem Grover's algorithm solves is so simple to understand -- given an oracle function that recognizes marked items, locate one of $M$ such marked items amongst $N$ unsorted items -- its classical time complexity $\mathcal{O}(N/M)$ is obvious, making the quantum speedup that much more conclusive.

Conceptually also, Grover's algorithm is compelling -- the iterative application of the oracle and initial state preparation rotates from a superposition of mostly unmarked states to a superposition of mostly marked states in just $\mathcal{O}(\sqrt{N/M})$ steps \cite{Aharonov1998}. This interpretation of Grover's algorithm as a rotation is very natural because the Grover iterate is a unitary operator. However, this same unitarity is also a weakness. Without knowing exactly how many marked items there are, there is no knowing when to stop the iteration! This leads to the souffl\'{e} problem \cite{Brassard1997}, in which iterating too little ``undercooks" the state, leaving mostly unmarked states, and iterating too much ``overcooks" the state, passing by the marked states and leaving us again with mostly unmarked states.

The most direct solution of the souffl\'{e} problem is to estimate $M$ by either using full-blown quantum counting \cite{Boyer1996, Brassard1998} or a trial-and-error scheme where iterates are applied an exponentially increasing number of times \cite{Boyer1996, Brassard2000}. Although scaling quantumly, these strategies are unappealing for search as they work best not by monotonically amplifying marked states, but rather by getting ``close enough" before resorting to classical random sampling.

An alternative approach, in line with what we advocate here, is to construct, either recursively or dissipatively, operators that avoid overcooking by {\it always} amplifying marked states. Such algorithms are known as fixed-point searches. For example, running Grover's $\pi/3$-algorithm \cite{Grover2005} or the comparable ancilla-algorithm \cite{Grover2006} longer can only ever improve its success probability. Yet, a steep price is paid for this monotonicity -- in both cases, the quadratic speedup of the original quantum search is lost.

This disappointing fact means that current fixed-point algorithms take time $\mathcal{O}(N/M)$ for small $M/N$, and their usefulness is relegated to large $M/N$, where they conveniently avoid overcooking, but where classical algorithms are also already successful. Several results \cite{Li2007, Toyama2009} improve the performance of fixed-point algorithms on wide ranges of $M/N$, but these algorithms are numerical and as such their time scaling cannot be assessed. Indeed, the $\pi/3$-algorithm was shown to be optimal in time \cite{Chakraborty2005}, ostensibly proving it impossible to find a search algorithm that both avoids the souffl\'{e} problem and provides a quantum advantage.

Nevertheless, here we present a fixed-point search algorithm, which, amazingly, achieves both goals -- our search procedure cannot be overcooked and also achieves optimal time scaling, a quadratic advantage over classical unordered search. We sidestep the conditions of the impossibility proof by requiring not that the error monotonically improve as in the $\pi/3$-algorithm, but that the error become bounded by a tunable parameter $\delta$ over an ever widening range of $M/N$ as our algorithm is run longer. The polynomial method \cite{Beals2001} is typically used to prove lower bounds on quantum query complexities; however, we instead use the fact that the success probability is a polynomial to adjust the phases of Grover's reflection operators \cite{Long1999, Hoyer2000} and effect an optimal output polynomial with bounded error $\delta$. In fact, our algorithm becomes the $\pi/3$-algorithm and Grover's original search algorithm in the special cases of $\delta=0$ and $\delta=1$, respectively. 

\begin{figure}
\centering
\includegraphics[width=\columnwidth]{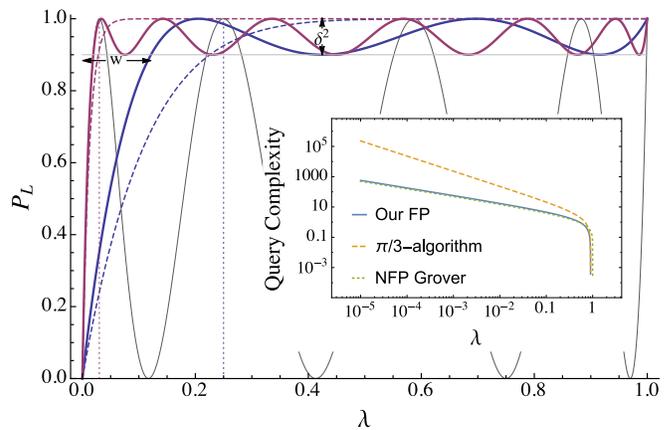}
\caption{A comparison of search algorithms, plotting the overlap $P_L$ of the target state with the output state versus the overlap $\lambda$ of the target state with the initial state. We weigh our fixed-point (FP) algorithm (thick solid) against the $\pi/3$-algorithm (dashed) for the task of achieving output success probability $P_L$ greater than $1-\delta^2=0.9$ for all $\lambda>\lambda_0$. The query complexity of the algorithms vary based on $\lambda_0$ (dotted vertical lines). For $\lambda_0=0.25$ (blue), our algorithm makes $4$ queries while the $\pi/3$-algorithm makes $8$. For $\lambda_0=0.03$ (red), our algorithm makes $12$ queries while the $\pi/3$-algorithm makes $80$. For comparison, also shown is Grover's non-fixed-point (NFP) search with 8 queries (thin black). The width and error for our 4-query algorithm are labeled $w$ and $\delta$, respectively. (Inset) We plot the query complexity against $\lambda$ for our algorithm with $\delta^2=0.1$ (solid), the $\pi/3$-algorithm (dashed), and non-fixed-point Grover's (dotted). While our FP algorthm and Grover's NFP algorithm scale as $L\sim1/\sqrt{\lambda}$, the $\pi/3$-algorithm scales as $L\sim1/\lambda$.}
\label{fig1}
\end{figure}

Our results apply just as cleanly, and more generally, to amplitude amplification \cite{Brassard2000}, so we proceed in that framework. We are given a unitary operator $A$ that prepares the initial state $\ket{s}=A\ket{0}^{\otimes n}$. From $\ket{s}$, we would like to extract the target state $\ket{T}$ with success probability $P_L\ge1-\delta^2$, where the overlap $\langle T|s\rangle=\sqrt{\lambda}e^{i\xi}$ is not zero and $\delta\in[0,1]$ is given. To do so, we are provided with the oracle $U$ which flips an ancilla qubit when fed the target state. That is, $U\ket{T}\ket{b}=\ket{T}\ket{b\oplus1}$ and $U\ket{\overline{T}}\ket{b}=\ket{\overline{T}}\ket{b}$ for $\langle\overline{T}|T\rangle=0$. Below, we show how to solve this problem and extract $\ket{T}$ by performing on $\ket{s}$ a quantum circuit $\mathcal{S}_L$ consisting of $A, A^\dag, U$, and efficiently implementable $n$-qubit gates, such that 
\begin{equation}\label{success_prob}
P_L=|\bra{T}\mathcal{S}_L\ket{s}|^2 = 1 - \delta^2\text{\space}T_L\left(T_{1/L}(1/\delta)\sqrt{1-\lambda}\right)^2.
\end{equation}
Here $T_L(x)=\cos(L\cos^{-1}(x))$ is the $L^{\text{th}}$ Chebyshev polynomial of the first kind \cite{Rivlin1990} and $L-1$ is the query complexity: the number of times $U$ is applied in the circuit $\mathcal{S}_L$. Furthermore, we will construct $\mathcal{S}_L$ for any odd integer $L\ge1$ and any $\delta$. Some examples of $P_L$ and a comparison to the $\pi/3$-algorithm are shown in Fig.~\ref{fig1}.

Assuming for now the existence of $\mathcal{S}_L$ -- its construction will be given later -- we can already see that the success probability $P_L$ possesses both the fixed point property and optimal query complexity. First, note that as long as $|T_{1/L}(1/\delta)|\sqrt{1-\lambda}\le 1$, the fact that $|T_L(x)|\le1$ for $|x|\le1$ implies $P_L\ge 1-\delta^2$. Therefore, for all $\lambda\ge w=1-T_{1/L}(1/\delta)^{-2}$, the probability $P_L$ meets our error tolerance. For large $L$ and small $\delta$, this width $w$ can be approximated as
\begin{equation}
w\approx\left(\frac{\log(2/\delta)}{L}\right)^2.
\end{equation}
This equation demonstrates the fixed-point property -- as $L$ increases, $w$ decreases, and we achieve success probability $P_L\ge1-\delta^2$ over an ever increasing range of $\lambda$. Equivalently, this means we cannot overcook the state, because if a sequence $\mathcal{S}_L$ achieves bounded error at $\lambda$, then so does $\mathcal{S}_{L'}$ for any $L'>L$. Second, note that to ensure the probability is bounded we must choose $L$ such that $w\le\lambda$. That is, for $\delta>0$,
\begin{equation}\label{complexity}
L\ge\frac{\log(2/\delta)}{\sqrt{\lambda}}.
\end{equation}
Thus, query complexity goes as $L=\mathcal{O}\left(\log(2/\delta)\frac{1}{\sqrt{\lambda}}\right)$ for our algorithm, achieving, for amplitude amplification, the best possible scaling in $\lambda$ \cite{Brassard2000}. See also Fig.~\ref{fig1} (inset).

Having seen two defining attributes, the fixed-point property and optimality, of the success probability from Eq.~\eqref{success_prob}, let us now create it using the operators provided: the state preparation $A$ and oracle $U$. This problem simplifies when interpreted in the two-dimensional subspace $\mathcal{T}$ spanned by $\ket{s}$ and $\ket{T}$ rather than in the full $2^n$-dimensional Hilbert space of all $n$ qubits. First, define $\ket{t}=e^{-i\xi}\ket{T}$ and $\ket{\overline{t}}=\big(\ket{s}-\langle t|s\rangle\ket{t}\big)/\sqrt{1-\lambda}$, so that
\begin{equation}\label{initial_state}
\ket{s}=\sqrt{1-\lambda}\ket{\overline{t}}+\sqrt{\lambda}\ket{t}=\left(\begin{array}{c}\sqrt{1-\lambda}\\\sqrt{\lambda}\end{array}\right).
\end{equation}
The matrix notation comes from the definitions $\ket{t}=\left(\begin{smallmatrix}0\\1\end{smallmatrix}\right)$ and $\ket{\overline{t}}=\left(\begin{smallmatrix}1\\0\end{smallmatrix}\right)$. The location of $\ket{s}$ on the Bloch sphere is in the XZ-plane at an angle $\phi$ from the north pole, where $\phi\in[0,\pi]$ is defined by $\sin(\phi/2)=\sqrt{\lambda}$. Our goal of achieving the $P_L$ of Eq.~\eqref{success_prob} is equivalently expressed as constructing, up to a global phase, the Chebyshev state
\begin{equation}\label{chebyshev_state}
\ket{C_L}=\sqrt{1-P_L}\ket{\overline{t}}+\sqrt{P_L}e^{i\chi}\ket{t}=\left(\begin{array}{c}\sqrt{1-P_L}\\\sqrt{P_L}e^{i\chi}\end{array}\right)
\end{equation}
for some relative phase $\chi$. For large enough $\lambda$, the Chebyshev state lies near the south pole of the Bloch sphere.

\begin{figure*}
\centering
\includegraphics[width=\textwidth]{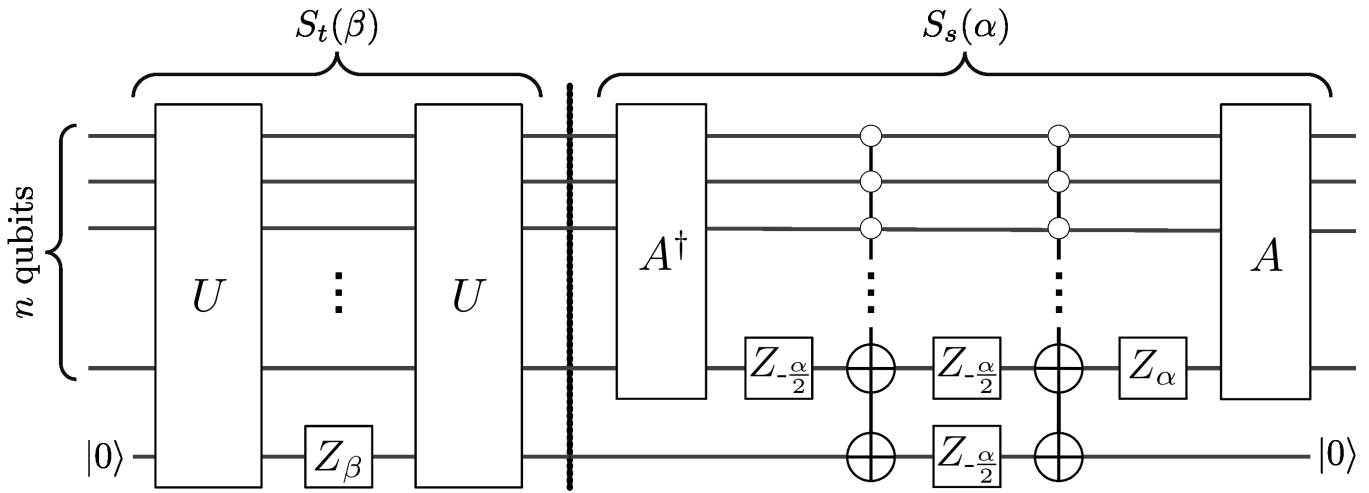}
\caption{We provide a circuit for performing the generalized Grover iterate $G(\alpha,\beta)$ up to a global phase. Here, $Z_\theta:=R_0(\theta)$ represents a rotation about the z-axis by angle $\theta$. The first part of the circuit, before the dotted line, performs $e^{-i\beta/2}S_t(\beta)$ and the second part performs $S_s(\alpha)$. One ancilla bit initialized as $\ket{0}$ is required for both parts, but can be reused. The multiply-controlled NOT gates in the $S_s(\alpha)$ circuit do not pose a substantial overhead -- they can be implemented with $\mathcal{O}(n^2)$ single qubit and CNOT gates \cite{Saeedi2013} or $\mathcal{O}(n)$ such gates and $\mathcal{O}(n)$ ancillas \cite{Nielsen2004}.}
\label{fig2}
\end{figure*}

Similarly, Grover's reflection operators can be interpreted as $\text{SU}(2)$ unitaries acting on $\mathcal{T}$. As in previous work \cite{Long1999, Hoyer2000}, we add arbitrary phases to the reflections to define generalized reflections. In Fig.~\ref{fig2} we show explicitly how to implement these generalized reflections using $A$, $U$, and efficiently implementable $n$-qubit operations. Their $\text{SU}(2)$ representations are
\begin{align}\label{generalized_reflection_s}
S_s(\alpha)&=I-(1-e^{-i\alpha})\ket{s}\bra{s}\\\nonumber
&=\left(\begin{array}{cc}1-(1-e^{-i\alpha})\overline{\lambda}&-(1-e^{-i\alpha})\sqrt{\lambda\overline{\lambda}}\\-(1-e^{-i\alpha})\sqrt{\lambda\overline{\lambda}}&1-(1-e^{-i\alpha})\lambda\end{array}\right),\\\label{generalized_reflection_t}
S_t(\beta)&=I-(1-e^{i\beta})\ket{t}\bra{t}=\left(\begin{array}{cc}1&0\\0&e^{i\beta}\end{array}\right),
\end{align}
where $\overline{\lambda}=1-\lambda$. The product of the reflection operators is often called the Grover iterate $G(\alpha,\beta)=-S_s(\alpha)S_t(\beta)$. The original Grover iterate \cite{Grover1996} used $\alpha=\pm\pi$ and $\beta=\pm\pi$. 

The generalized reflection operators are also expressible as rotations on the Bloch sphere. Defining $R_\varphi(\theta)=\exp\left(-i\frac12\theta(\cos(\varphi)Z+\sin(\varphi)X)\right)$ for Pauli operators $X$ and $Z$, we find
\begin{align}\label{as_rotation_s}
S_s(\alpha)&=e^{-i\alpha/2}R_\phi(\alpha)\\\label{as_rotation_t}
S_t(\beta)&=e^{i\beta/2}R_0(\beta).
\end{align}
When $\alpha=\pm\pi$ and $\beta=\pm\pi$, these rotations map the XZ-plane to the XZ-plane, reproducing the $\text{O}(1)$ rotation picture of Grover's original non-fixed-point algorithm \cite{Aharonov1998}. 

Yet, why limit ourselves to $O(1)$ when, by using general phases $\alpha$ and $\beta$, we can access the whole of $\text{SU}(2)$? To that end, we consider a sequence of $l$ generalized Grover iterates. Since each generalized Grover iterate contains two queries to $U$, such a sequence would have query complexity $L-1=2l$. We thus set out to find, for any $\lambda>0$, phases $\alpha_j$ and $\beta_j$ such that the sequence
\begin{equation}\label{sequence}
\mathcal{S}_L=G(\alpha_l,\beta_l)\dots G(\alpha_1,\beta_1)=\prod_{j=1}^lG(\alpha_j,\beta_j)
\end{equation}
attains success probability $P_L$ by preparing, up to a global phase, the Chebyshev state: $|\bra{C_L}\mathcal{S}_L\ket{s}|=1$.

Indeed, such phases exist for all $l$ and all $\delta\in[0,1]$, and, moreover, they may be given in very simple analytical forms. For all $j=1,2,\dots,l$, we have
\begin{align}\label{phase_solution}
\alpha_j=-\beta_{l-j+1}=2\cot^{-1}\left(\tan(2\pi j/L)\sqrt{1-\gamma^2}\right),
\end{align}
where $L=2l+1$ as before and $\gamma^{-1}=T_{1/L}\left(1/\delta\right)$. Notice Grover's non-fixed-point search is subsumed by this solution -- if $\delta=1$, then $\alpha_j=\pm\pi$ and $\beta_j=\pm\pi$ for all $j$, values that we saw above give Grover's original non-fixed-point algorithm \cite{Grover1996}. Thus, when $\delta=1$, our algorithm is exactly Grover's search. 

The proof that Eq.~\eqref{phase_solution} implies Eq.~\eqref{success_prob} begins by rearranging $\mathcal{S}_L$. Let $A_\zeta=\exp(-i\frac12\phi(\cos(\zeta)X+\sin(\zeta)Y))$. With this definition, the state preparation operator is $A=A_{\pi/2}$. Also note the identities $R_{\phi}(\alpha)=A_{\pi/2}R_0(\alpha)A_{-\pi/2}$ and $A_{\alpha+\beta}=R_0(\beta)A_\alpha R_0(-\beta)$. Then, using Eqs.~(\ref{as_rotation_s}-\ref{as_rotation_t}), we find, up to a global phase, that
\begin{equation}\label{amp_err_seq}
\mathcal{S}_L\ket{s}\sim R_0(\zeta_1)\left(A_{\zeta_L}\dots A_{\zeta_2}A_{\zeta_1}\right)R_0(-\zeta_1)\ket{0}.
\end{equation}
Here the phases $\zeta_k=\zeta_{L-k+1}$ are palindromic, a consequence of the phase matching $\alpha_j=-\beta_{l-j+1}$. With $\alpha_j$ defined by Eq.~\eqref{phase_solution}, all $\zeta_k$ can be found recursively using $\zeta_{l+1}=(-1)^l\pi/2$ and
\begin{equation}\label{amp_err_phases}
\zeta_{k+1}-\zeta_k=(-1)^k\pi-2\cot^{-1}\left(\tan(k\pi/L)\sqrt{1-\gamma^2}\right)
\end{equation}
for all $k=1,\dots,L-1$.

From Eq.~\eqref{amp_err_seq}, we set up a recurrence relation to study the amplitude in states $\ket{t}$ and $\ket{\overline{t}}$ after each application of $A_\zeta$. That is, we let $(a_0,b_0)=(1,0)$ and for $h=1,\dots,L$ define $a_h$ and $b_h$ by the matrix equation
\begin{equation}\label{matrix_recursion}
\left(\begin{array}{c}a_h\\b_h\sin(\phi/2)\end{array}\right)=A_{\zeta_h}\left(\begin{array}{c}a_{h-1}\\b_{h-1}\sin(\phi/2)\end{array}\right).
\end{equation}
Letting $x=\cos(\phi/2)$, we can decouple this recurrence by defining $b'_h=-xa_h-i\sqrt{1-x^2}e^{-i\zeta_h}b_h$. Rearranging Eq.~\eqref{matrix_recursion}, we find $b'_h=-a_{h-1}$ and
\begin{equation}
a_h=x(1+e^{-i(\zeta_h-\zeta_{h-1})})a_{h-1}-e^{-i(\zeta_h-\zeta_{h-1})}a_{h-2},
\end{equation}
for $h=2,\dots,L$ with initial values $a_0=1$ and $a_1=x$. This recurrence is strikingly similar to that defining the Chebyshev polynomials: $T_n(x)=2xT_{n-1}(x)-T_{n-2}(x)$. Indeed, using Eq.~\eqref{amp_err_phases}, the Chebyshev recurrence is exactly recovered when $\gamma=\delta=1$. For other values of $\gamma$, the complex, degree-$h$ polynomials $a_h^{(\gamma)}(x)$ generalize the Chebyshev polynomials. In fact, it can be shown using combinatorial arguments analogous to those in \cite{Benjamin2009} that $a^{(\gamma)}_L(x)=\frac{T_L(x/\gamma)}{T_L(1/\gamma)}$. Since $T_L(1/\gamma)=1/\delta$ and $P_L=1-|a^{(\gamma)}_L(x)|^2$, this completes the proof of Eq.~\eqref{success_prob}.

While the solutions in Eq.~\eqref{phase_solution} are extremely simple to express, there are other solutions. Indeed, solutions of small length $l$ and large width $w$ can be combined to create solutions of larger length and smaller width through a process we call nesting. The general idea of nesting is that, within a sequence $\mathcal{S}_{L_2}$, the state preparation $A$ can be replaced by another sequence $\mathcal{S}_{L_1}A$ to recursively narrow the region of high failure probability. An intuition for this recursion can be noted in the similarity of Eq.~\eqref{initial_state} and Eq.~\eqref{chebyshev_state}. Nesting is similar to concatenation in composite pulse sequence literature \cite{Jones2013} and has already been employed in special cases of fixed-point search \cite{Grover2005}.

Although nesting would work to widen any fixed-point sequence (those found in \cite{Li2007,Toyama2009}, for instance), with our sequences using phases from Eq.~\eqref{phase_solution}, nesting neatly preserves the form of the success probability $P_L$. For notational convenience let us denote by $\mathcal{S}_L(B)$ a sequence of generalized Grover iterates as in Eq.~\eqref{sequence} that uses $BA$ in place of the state preparation operator $A$. For instance, with $I$ the identity operator, we know
\begin{equation}
\mathcal{S}_{L_1}\left(I\right)\ket{s}=\sqrt{1-P_{L_1}(\lambda)}\ket{\overline{t}}+\sqrt{P_{L_1}(\lambda)}e^{i\chi_1}\ket{t},
\end{equation}
where we have made explicit the dependence of $P_L$ from Eq.~\eqref{success_prob} on $\lambda$. By the same logic,
\begin{align}\label{nested_sequence}
\mathcal{S}_{L_2}\left(\mathcal{S}_{L_1}(I)\right)\mathcal{S}_{L_1}(I)\ket{s}&=\sqrt{1-P_{L_2}\left(P_{L_1}(\lambda)\right)}\ket{\overline{t}}\\\nonumber&+\sqrt{P_{L_2}\left(P_{L_1}(\lambda)\right)}e^{i(\chi_1+\chi_2)}\ket{t}.
\end{align}
Consider $P_{L_2}\left(P_{L_1}(\lambda)\right)$ and say that we choose the error bound for sequence 1 to be $\delta_1=\left(T_{1/L_2}\left[1/\delta\right]\right)^{-1}$ and that for sequence 2 to be $\delta_2=\delta$. Using the semi-group property of the Chebyshev polynomials, $T_p\left(T_q\left(x\right)\right)=T_{pq}(x)$, simple algebra yields
\begin{align}
P_{L_2}\left(P_{L_1}(\lambda,\delta_1),\delta_2\right)=P_{L_1L_2}(\lambda,\delta),
\end{align}
where we have further explicated the dependence of $P_L$ from Eq.~\eqref{success_prob} on its error bound $\delta$.

Therefore, as a result of nesting we can combine sequences of complexities $L_1$ and $L_2$ to obtain a sequence of complexity $L_1L_2$. In terms of Grover iterations, sequences with $l_1$ and $l_2$ iterations can be combined into one with $l=l_1+2l_1l_2+l_2$ iterations. If the phase angles of the component sequences are denoted $\alpha^{(1)}_j$ and $\alpha^{(2)}_j$ then the nested sequence has phase angles
\begin{equation}
\alpha^{(1,2)}_j=\begin{cases}\alpha^{(1)}_h&j\equiv h\text{\space}(\text{mod}\text{\space}L_1)\\-\alpha^{(1)}_h&j\equiv -h\text{\space}(\text{mod}\text{\space}L_1)\\\alpha^{(2)}_k&j=kL_1\end{cases}
\end{equation}
where $h\in\{1,2,\dots,l_1\}$ and $k\in\{1,2,\dots,l_2\}$. The accompanying phase angles $\beta^{(1,2)}_j$ can be taken to be phase matched, $\beta^{(1,2)}_j=-\alpha^{(1,2)}_{l-j+1}$.

With nesting, we can see that the $\pi/3$-algorithm \cite{Grover2005} is a special case of ours. From Eq.~\eqref{phase_solution}, note that our $l=1$ sequence with $\delta=0$ has phases $-\alpha_1=\beta_1=\pi/3$ and nesting it with itself gives exactly the $\pi/3$-algorithm. The query complexity argument represented by Eq.~\eqref{complexity} breaks down when $\delta=0$. In fact, the complexity of the $\pi/3$-algorithm scales classically as $\mathcal{O}\left(\frac{1}{\lambda}\right)$ \cite{Grover2005,Grover2006}.

A strong argument for using nesting, even though explicit solutions at all lengths are available in Eq.~\eqref{phase_solution}, is that it lends our algorithm a nice property: adaptability. At the end of any sequence $\mathcal{S}_{L_1}$, we can choose to keep the result, the Chebyshev state $\ket{C_{L_1}}$, or enhance it further to the Chebyshev state $\ket{C_{L_1L_2}}$ for any odd $L_2$. So, conveniently, sequences can be extended without restarting the algorithm from the initial state $\ket{s}$. This works because $\mathcal{S}_{L_1}$ is a prefix of the nested sequence in Eq.~\eqref{nested_sequence}. This is not something the phases with the form in Eq.~\eqref{phase_solution} allow as written, since they are prefix-free.

Our fixed-point algorithm can be used as a subroutine in any scenario where amplitude amplification or Grover's search is used \cite{Ambainis2004}, including quantum rejection sampling \cite{Ozols2013}, optimum finding \cite{Durr1996, Aaronson2006}, and collision problems \cite{Brassard1997b}. The obvious advantage of our approach over Grover's original algorithm is that there is no need to hunt for the correct number of iterations as in \cite{Boyer1996}, and this consequently eliminates the need to ever remake the initial state and restart the algorithm. Ideally, no measurements at all are required if $\delta$ and $L$ are chosen so the error of any amplitude amplification step will not significantly affect the error of the larger algorithm of which it is a part. Thus, our fixed-point amplitude amplification could make such algorithms completely coherent. 

An interesting direction for future work is relating quantum search to filters. In fact, the Dolph-Chebyshev function in Eq.~\eqref{success_prob} is one of many frequency filters studied in electronics \cite{Harris1978}. For our purposes, the Dolph-Chebyshev function guarantees the maximum range of $\lambda$ over which the bound $P_L\ge1-\delta^2$ can be satisfied by a polynomial of degree $L$ \cite{Dolph1946}. Moreover, since the probability of success is guaranteed to be polynomial in $\lambda$ and its degree is proportional to the number of queries made \cite{Beals2001}, we can also see this range is the maximum achievable with $\mathcal{O}(L)$ queries.

Our algorithm is also easily modified to avoid the target state -- simply using $\alpha_j$ from Eq.~\eqref{phase_solution}, but with $\beta_{l-j+1}=\alpha_j$ instead, will amplify the component of $\ket{s}$ that lies perpendicular to $\ket{T}$, so that $|\bra{\overline{T}}\mathcal{S}_L\ket{s}|^2=P_L$. Using this insight, it is tempting for instance to consider ``trapping" magic states \cite{Bravyi2005} by repelling a slightly non-stabilizer state from all the stabilizer states nearby.

Similar to the $\pi/3$-algorithm \cite{Reichardt2005}, our sequences also have application to the correction of single qubit errors, as suggested by Eq.~\eqref{amp_err_seq}. For instance, if a perfect bit-flip $X$ is desired, but only another non-identity operation $A\in\text{SU}(2)$, its inverse $A^\dag$, and perfect Z-rotations are available, then, still, the operator $X$ can be implemented with high-fidelity. Such a situation is reality for some experiments -- for example, those with amplitude errors \cite{Merrill2012}.

We gratefully acknowledge funding from NSF RQCC project \#1111337 and the ARO Quantum Algorithms Program. TJY acknowledges the support of the NSF iQuISE IGERT program.

\bibliography{OptimalFixedPoint}

\end{document}